\documentstyle[twocolumn]{mn}
\newif\ifAMStwofonts

\ifoldfss
  \ifCUPmtlplainloaded \else
    \NewTextAlphabet{textbfit} {cmbxti10} {}
    \NewTextAlphabet{textbfss} {cmssbx10} {}
    \NewMathAlphabet{mathbfit} {cmbxti10} {} 
    \NewMathAlphabet{mathbfss} {cmssbx10} {} 
  \fi
  \ifAMStwofonts
    \ifCUPmtlplainloaded \else
      \NewSymbolFont{upmath} {eurm10}
      \NewSymbolFont{AMSa} {msam10}
      \NewMathSymbol{\upi}     {0}{upmath}{19}
      \NewMathSymbol{\umu}     {0}{upmath}{16}
      \NewMathSymbol{\upartial}{0}{upmath}{40}
      \NewMathSymbol{\leqslant}{3}{AMSa}{36}
      \NewMathSymbol{\geqslant}{3}{AMSa}{3E}

       \let\le=\leqslant
       
    \fi
  \fi
\fi 

\ifnfssone
  \newmathalphabet{\mathit}
  \addtoversion{normal}{\mathit}{cmr}{m}{it}
  \addtoversion{bold}{\mathit}{cmr}{bx}{it}
  \newmathalphabet{\mathbfit} 
  \addtoversion{normal}{\mathbfit}{cmr}{bx}{it}
  \addtoversion{bold}{\mathbfit}{cmr}{bx}{it}
  \newmathalphabet{\mathbfss} 
  \addtoversion{normal}{\mathbfss}{cmss}{bx}{n}
  \addtoversion{bold}{\mathbfss}{cmss}{bx}{n}
  \ifAMStwofonts
    \ifCUPmtlplainloaded \else
      \UseAMStwoboldmath
      \makeatletter
      \new@mathgroup\upmath@group
      \define@mathgroup\mv@normal\upmath@group{eur}{m}{n}
      \define@mathgroup\mv@bold\upmath@group{eur}{b}{n}
      \edef\UPM{\hexnumber\upmath@group}
      \new@mathgroup\amsa@group
      \define@mathgroup\mv@normal\amsa@group{msa}{m}{n}
      \define@mathgroup\mv@bold\amsa@group{msa}{m}{n}
      \edef\AMSa{\hexnumber\amsa@group}
      \makeatother
      \mathchardef\upi="0\UPM19
      \mathchardef\umu="0\UPM16
      \mathchardef\upartial="0\UPM40
      \mathchardef\leqslant="3\AMSa36
      \mathchardef\geqslant="3\AMSa3E

       \let\le=\leqslant

    \fi
  \fi
\fi 

\ifnfsstwo
  \DeclareMathAlphabet{\mathbfit}{OT1}{cmr}{bx}{it}
  \SetMathAlphabet\mathbfit{bold}{OT1}{cmr}{bx}{it}
  \DeclareMathAlphabet{\mathbfss}{OT1}{cmss}{bx}{n}
  \SetMathAlphabet\mathbfss{bold}{OT1}{cmss}{bx}{n}
  \ifAMStwofonts
    \ifCUPmtlplainloaded \else
      \DeclareSymbolFont{UPM}{U}{eur}{m}{n}
      \SetSymbolFont{UPM}{bold}{U}{eur}{b}{n}
      \DeclareSymbolFont{AMSa}{U}{msa}{m}{n}
      \DeclareMathSymbol{\upi}{0}{UPM}{"19}
      \DeclareMathSymbol{\umu}{0}{UPM}{"16}
      \DeclareMathSymbol{\upartial}{0}{UPM}{"40}
      \DeclareMathSymbol{\leqslant}{3}{AMSa}{"36}
      \DeclareMathSymbol{\geqslant}{3}{AMSa}{"3E}

       \let\le=\leqslant

    \fi
  \fi
\fi 

\ifCUPmtlplainloaded \else
  \ifAMStwofonts \else 
    \def\upi{\pi}
    \def\umu{\mu}
    \def\upartial{\partial}
  \fi
\fi

\input epsf
\title[Multi-fluid   Magnetohydrodynamics]{A   Numerical  Scheme   for
Multi-fluid Magnetohydrodynamics}

\author[S.~A.~E.~G.  Falle ]{S.~A.~E.~G.~Falle\\Department of Applied
Mathematics, University of Leeds, Leeds LS2 9JT, UK.}

\begin{document}

\maketitle

\begin{abstract}
This paper describes a  numerical scheme for multi-fluid hydrodynamics
in the  limit of  small mass densities  of the charged  particles. The
inertia of the charged particles can then be neglected, which makes it
possible to  write an evolution  equation for the magnetic  field that
can  be solved  using  an  implicit scheme.   This  avoids the  severe
restriction on the stable timestep  that would otherwise arise at high
resolution, or  when the  Hall effect is  large. Numerical  tests show
that  the  scheme  can   accurately  model  steady  multi-fluid  shock
structures both with and  without sub-shocks. Although the emphasis is
on  shocks in molecular  clouds, a  multi-dimensional version  of this
code could  be applied to  any Astrophysical flow in  which ambi-polar
diffusion or the Hall effect, or both play a significant role.
\end{abstract}

\begin{keywords}
ISM: MHD - shock waves - dust, numerical methods, molecules.
\end{keywords}

\section{Introduction}

In dense molecular clouds, the  density of charged particles can be so
low that  the scale on  which ambi-polar resistivity  becomes important
cannot be  assumed to  be neglibly small.  The most obvious  effect of
this is that it allows the existence of shock structures that are much
thicker than those  determined by viscous effects (see  e.g. Draine \&
McKee 1993), but the enhanced  magnetic diffusion also plays a role in
the large scale dynamics of such clouds (see e.g. Mouschovias 1991).

The structure of shocks in  which the dissipation is due to ambi-polar
resistivity rather  than viscosity has been  studied extensively (e.g.
Mullan  1971; Draine 1980;  Flower, Pineau  des For\^ets  \& Hartquist
1985; ; Draine 1986; Wardle  \& Draine 1987; Chernoff 1987; Roberge \&
Draine  1990;  Pilipp \&  Hartquist  1994;  Wardle  1998).  There  are
several reasons for this, of  which perhaps the most important is that
the flow time  through these shocks is long  enough for radiation from
molecules  to maintain the  gas at  a low  temperature even  for shock
speeds as  high as 50 km  s$^{-1}$.  Indeed, Wardle  (1998) points out
that this process is so effective that such shocks are responsible for
much of the  infra-red H$_2$ and CO line  emission in molecular clouds
(Draine \& Roberge 1982; Chernoff,  McKee \& Hollerbach 1982; Smith \&
Brand  1990;  Smith, Brand  \&  Moorhouse  1991;  Chrysostomou et  al.
1997).  The heating due to the currents also raises the temperature of
the molecular  gas to the point  where a number  of important chemical
reactions proceed  at a significant rate (Draine,  Roberge \& Dalgarno
1983;  Flower,  Pineau des  For\^ets  \&  Hartquist  1985; Pineau  des
For\^ets, Flower  Hartquist \& Dalgarno 1986; Draine  \& Katz 1986a,b;
Kaufman \& Neufeld 1996a,b).

Although the obvious  way to determine the steady  shock structures is
to solve the steady equations directly, this is not a simple matter in
the  general  case. There  is  no  great  difficulty when  the  steady
solution can be  obtained by integrating through the  structure in the
appropriate direction  (e.g. Wardle \&  Draine 1987) and one  can also
deal with solutions containing sub-shocks if the system reduces to two
differential equations (Chernoff 1987). However, these are all special
cases and the general case of oblique shocks with cooling and chemical
reactions  is a  two point  boundary value  problem that  can  only be
solved by a  relaxation method such as that used  by Draine (1980). In
that case  it is much simpler to  use a time dependent  code to obtain
the steady solution.  This also has the advantage that  it can be used
for unsteady problems in which these effects are important.

A number  of numerical schemes  for the time dependent  equations have
been devised  (T\'oth 1994; Smith  \& Mac Low  1997; Mac Low  \& Smith
1997; Stone 1997; Chieze, Pineau de For\^ets \& Flower 1998; Ciolek \&
Roberge 2002),  but, with the  exception of Ciolek \&  Roberge (2002),
these  all assume that  all the  charged particles  have a  large Hall
parameter and  are therefore tied  to the magnetic field.   The system
can then be modeled as three fluids: a neutral fluid, an ion fluid and
an  electron  fluid.   The  ions  and  electrons  may  have  different
temperatures,  but the  magnetic field  forces them  to have  the same
velocity, which  can, however, be  different from that of  the neutral
fluid. 

In molecular clouds,  the assumption that the Hall  parameter is large
is  valid for  electrons, ions  and  small grains  such as  polycyclic
aromatic hydrocarbons  (PAH), but not for grains  whose radius, $a_g$,
is larger  than about  $10^{-5}$ cm.  Pilipp  \& Hartquist  (1994) and
Wardle  (1998)  have  shown  that  a  significant  charge  density  of
particles for which the Hall parameter is $O(1)$ has a profound effect
on the shock structures.  In  particular, such particles induce a Hall
resistivity which  can lead to substantial rotation  of the transverse
field within the shock structure.   For this reason, Ciolek \& Roberge
(2002) devised an algorithm for  time dependent equations that include
a significant charge density of  particles whose Hall parameter is not
large enough for  them to be tied to the field.   However, as we shall
see, the  stable timestep for  their algorithm can become  very small,
especially  if there is  a significant  Hall resistivity.   This paper
describes an algorithm that does not suffer from this restriction.

Although  the emphasis  here is  on  shocks in  molecular clouds,  the
algorithm is quite  general and is efficient for  any problem in which
either  ambi-polar   diffusion  or  the  Hall  effect,   or  both  are
important. In  particular, although ambi-polar  diffusion in molecular
clouds is due to drift  between neutrals and charged particles, it can
also arise due  to drift between electrons and ions  in plasmas with a
small fraction of  neutrals if the density is high  enough for the ion
Hall parameter to be $O(1)$ (e.g. Sano \& Stone 2002).
 
\section{Multi-fluid Equations}

Consider a  set of  $N$ fluids for  which the generic  one dimensional
equations are ($i = 1 \cdots N$)

\medskip

\begin{equation}
{{\partial \rho_i} \over {\partial  t}} + {{\partial \rho_i u_i} \over
{\partial x}} = \sum_{j\not=i}^N s_{ij},
\label{coneq}
\end{equation}

\begin{equation}
\begin{array}{lcl}
\displaystyle{{{\partial  \rho_i  {\bf  q}_i}  \over {\partial  t}}  +
{\partial \over  {\partial x}} (\rho_i  u_i {\bf q}_i +  p_i \hat{{\bf
{\i}}})}  & =  & \displaystyle{\alpha_i  \rho_i ({\bf  E} +  {\bf q}_i
\wedge {\bf B})} \\
 & & \displaystyle{+ \sum_{j\not=i}^N {{\bf f}_{ij}}},\\
\end{array}
\label{momeq}
\end{equation}

\begin{equation}
\begin{array}{lcl}
\displaystyle{{{\partial  e_i} \over {\partial  t}} +  {\partial \over
{\partial x}}  [u_i (e_i +  p_i +  {1 \over 2}  \rho_i q_i^2)]} &  = &
\displaystyle{H_i + \sum_{j\not=i}^N {G_{ij}}} \\
\multicolumn{3}{r}{\displaystyle{+  \alpha_i  \rho_i  {\bf q}_i  \cdot
({\bf E} + {\bf q}_i \wedge {\bf B})}}.\\
\end{array}
\label{eneq}
\end{equation}

\noindent
Here ${\bf q}_i = (u_i,  v_i, w_i)$ are the velocities, $\alpha_i$ the
charge to mass ratios, $e_i$ the total energies and $\hat{{\bf {\i}}}$
the unit vector in the x direction. We have

\[
e_i = \rho_i (U_i + {1 \over 2} {\bf q}_i^2)
\]

\noindent
where $U_i$  is the  internal energy  per unit mass  of fluid  $i$. We
shall assume that

\[
U_i = W_i + {p_i \over {\rho_i (\gamma_i - 1)}}
\]

\noindent
where $W_i$ is  the energy associated with the  internal states of the
particles  of fluid $i$.  It can  include ionization  energy, chemical
binding  energy  etc.

The source terms are: $s_{ij}$ -- mass transfer rate from fluid $j$ to
fluid $i$; ${\bf f}_{ij}$ --  momentum transfer rate from fluid $j$ to
fluid $i$;  $G_{ij}$ -- energy transfer  rate from fluid  $j$ to fluid
$i$; $H_i$ -- external energy  source/sink for fluid $i$. Global mass,
energy  and momentum  conservation clearly  require that  $s_{ij}  = -
s_{ji}$,  ${\bf f}_{ij} =  - {\bf  f}_{ji}$ and  $G_{ij} =  - G_{ji}$.
Draine (1986) derives expressions for these terms, but for our purpose
it is sufficient to consider their general form.

The momentum transfer rate, ${\bf f}_{ij}$, is

\[
{\bf f}_{ij} = {\bf C}_{ij} + s_{ij} {\bf q}_j - s_{ji} {\bf q}_i,
\]

\noindent
where  ${\bf C}_{ij}$  describes collisions  between the  particles of
fluids $i$ and $j$.  Since  it involves binary collisions and tends to
equalise the velocities of the two fluids, it must be of the form

\[
{\bf C}_{ij} = \rho_i \rho_j K_{ij}(T_i, T_j, |{\bf q}_j - {\bf q}_i|)
({\bf q}_j - {\bf q}_i),
\]

\noindent
where  $T_i$ and $T_j$  are the  temperatures of  fluids $i$  and $j$.
Since  global momentum conservation  requires ${\bf  C}_{ij} =  - {\bf
C}_{ji}$, we must have ${\bf K}_{ij} = {\bf K}_{ji}$.

The energy transfer rate, $G_{ij}$, is

\[
G_{ij} =  {\bf q}_i  \cdot {\bf C}_{ij}  +s_{ij}  e_j - s_{ji}  e_i +
D_{ij},
\]

\noindent
where  $D_{ij}$  describes  the   effect  of  collisions  between  the
particles of  fluids $i$  and $j$.  Like  ${\bf C}_{ij}$,  it involves
binary  collisions and tends  to produce  equilibrium between  the two
fluids. It must therefore be of the form

\[
D_{ij} = \rho_i \rho_j L_{ij}(T_i, T_j, |{\bf q}_j - {\bf q}_i|).
\]

\noindent
with $L_{ij} = 0$ when $T_i = T_j$ and ${\bf q}_i = {\bf q}_j$. Global
energy conservation requires

\[
G_{ij} =  {\bf q}_i \cdot {\bf  C}_{ij} + D_{ij}  = - G_{ji} =  - {\bf
q}_j \cdot {\bf C}_{ij} + D_{ji}.
\]

The fields are determined from Maxwell's equations, which reduce to

\begin{equation}
{{\partial  B_x} \over  {\partial  t}} =  0,~~~~{{\partial B_y}  \over
{\partial  t}} =  {{\partial E_z}  \over  {\partial x}},~~~~{{\partial
B_z}  \over {\partial  t}} =  - {{\partial  E_y} \over  {\partial x}},
\label{Fara}
\end{equation}

\begin{equation}
{{\partial B_y} \over {\partial x}} = J_z,~~~~ {{\partial B_z} \over
{\partial x}} = - J_y
\label{Amp}
\end{equation}

\noindent
since the  displacement current may  be neglected. The  current, ${\bf
J}$, is given by

\begin{equation}
{\bf J} = \sum_{i=1}^N { \alpha_i \rho_i {\bf q}_i}.
\label{curr}
\end{equation}

\noindent
and we also have charge neutrality

\begin{equation}
\sum_{i=1}^N { \alpha_i \rho_i} = 0.
\label{charneut}
\end{equation}

\noindent
The units  for ${\bf E}$,  ${\bf B}$ and  ${\bf J}$ are such  that the
speed of light and the factor $4 \pi$ do not appear.

The above equations are extremely general and include those that other
authors have used to model shocks in molecular clouds as special cases
Despite  this  generality, they  do  assume  that  each fluid  can  be
described by the hydrodynamic approximation, which is only true if the
interactions between particles of the different fluids are much weaker
than those between particles of the same fluid.

In the  usual applications,  the mass is  dominated by a  fluid, $i=1$
say, consisting of neutral  particles, which obviously has zero charge
to mass  ratio. The other fluids are:  an electron fluid, $i  = 2$; an
ion fluid, $i  = 3$; fluids consisting of grains  of various types and
sizes,  $i  =   4  \cdots  N$.   The  validity   of  the  hydrodynamic
approximation for  systems consisting only of  neutrals, electrons and
ions has  been discussed  by Draine (1986)  and Chernoff  (1987). They
conclude that the hydrodynamic approximation is valid for the neutrals
and electrons,  both of which  should have a  Maxwellian distribution,
but this may not be true for the ions.

The grain  fluids are obviously  very different since they  consist of
particles with much larger masses than the other fluids. Their thermal
velocity  dispersion   is  therefore  small  compared   to  the  drift
velocities, which  means that all  the particles of a  particular size
and type have the  same velocity.  The hydrodynamic approximation with
zero pressure is therefore appropriate for the grain fluids.

Although  it is  possible to  construct a  numerical scheme  for these
equations as they stand, this would not be suitable for the conditions
in shocks in molecular clouds.  Consider the neutral momentum equation
and  suppose that  the  drift  velocities are  small  compared to  the
thermal velocity dispersion of the  neutrals, as they must be near the
upstream  or downstream  ends of  the shock  structure.  The collision
term is then approximately

\[
\sum_{i=2}^N   {{\bf   C}_{1i}}    =   \sum_{i=2}^N   {\rho_1   \rho_i
K_{1i}(T_1,T_i,0) ({\bf q}_i - {\bf q}_1)}.
\]

If $q_c$ is a typical velocity, such as the shock velocity relative to
the upstream fluid,  then a balance between this  term and the inertia
terms induces a length scale

\[
l_{c1} = {q_c \over {\sum_{i=2}^N {K_{1i}(T_1,T_i,0) \rho_i}}}.
\]

\noindent
It is clear that $l_{c1}$  determines the shock thickness.  Then since
$\rho_i \ll \rho_1$ for all $i  > 1$, the equivalent length scales for
the other fluids are

\[
l_{ci} = {q_c \over {K_{1i}(T_1,T_j,0) \rho_1}},
\]

\noindent
which  means that  $l_{ci} \ll  l_{c1}$.

The charged fluids are also  acted on by electromagnetic forces, which
induce a length scale

\[
l_{ei} = {{q_c} \over {\alpha_i B}},
\]

\noindent
which is the Larmor radius of particles with velocity $q_c$.

Provided  either  $l_{ci} \ll  l_{c1}$  or  $l_{ei}  \ll l_{c1}$,  the
inertia and pressure terms can be ignored in the momentum equation for
fluid $i$.  Since  this is generally true, the  momentum equations for
all fluids except the neutral fluid reduce to

\begin{equation}
{\alpha_i \rho_i ({\bf E} + {{{\bf q}_i} \over c}  \wedge {\bf B})} +
{\bf f}_{i1} = 0,
\label{redmom}
\end{equation}

\noindent
where the fact that $\rho_i \ll  \rho_1$ for all $i$ has allowed us to
ignore momentum  transfer from  all but the  neutral fluid.   The same
arguments  tell us  that in  this case  the energy  equations  for the
charged fluids reduce to

\begin{equation}
H_i +  G_{i1} + \alpha_i \rho_i {\bf  q}_i \cdot ({\bf E}  + {\bf q}_i
\wedge {\bf B})= 0
\label{reneqi}
\end{equation}

If $l_{ei}  \ll l_{c1}$,  then particles $i$  are closely tied  to the
field lines. It  is convenient to define a  Hall parameter, $\beta_i$,
for each charged fluid by

\[
\beta_i  =  {l_{ci}  \over  l_{ei}}  =  {{\alpha_i  B}  \over  {K_{1i}
\rho_1}}.
\]

\noindent
The ions and electrons have $\beta_i  \gg 1$ and are therefore tied to
the  field, whereas  the  grains  are not  since  they typically  have
$\beta_i \simeq 1$.

\section{Numerical Method}

Our method deals with the  same equations as Ciolek \& Roberge (2002),
but uses a somewhat different  approach, which, as we shall see, leads
to  a  more robust  and  efficient scheme.

\subsection{Equations for the Neutral Fluid}

Since ${\bf f}_{ij} = -  {\bf f}_{ji}$, the reduced momentum equations
for  the   charged  fluids,  (\ref{redmom}),   the  charge  neutrality
condition,  (\ref{charneut})  and  the  expression  for  the  current,
(\ref{curr}) give

\[
\sum_{i=2}^N  {{\bf  f}_{1i}}  =   -  \sum_{i=2}^N  {{\bf  f}_{i1}}  =
\sum_{i=2}^N {\alpha_i \rho_i ({\bf E}  + {\bf q}_i \wedge {\bf B})} =
{\bf J} \wedge {\bf B}.
\]

\noindent
The  momentum  equation for  the  neutral  fluid, (\ref{momeq}),  then
becomes

\begin{equation}
{{\partial  \rho_1 {\bf q}_1}  \over {\partial  t}} +  {\partial \over
{\partial  x}}  (\rho_1  u_1  {\bf  q}_1 +  p_1  \hat{{\bf  {\i}}})  =
\sum_{i=2}^N {{\bf f}_{1i}} = {\bf J} \wedge {\bf B}.
\label{avmomeq}
\end{equation}

Similarly, the reduced energy equations, (\ref{reneqi}), give

\[
\begin{array}{l}
\displaystyle{\sum_{i=2}^N {G_{1i}} = - \sum_{i=2}^N {G_{i1}} = \sum_{i=2}^N {[H_i +
\alpha_i \rho_i {\bf q}_i \cdot ({\bf E} + {\bf q}_i \wedge {\bf B})]}
} \\
 \\
~~~~~~~~~~~=\displaystyle{{\bf J} \cdot {\bf E} + \sum_{i=2}^N {[H_i +
\alpha_i \rho_i {\bf q}_i \cdot ({\bf q}_i \wedge {\bf B})]}}.
\end{array}
\]

\noindent
The energy equation for the neutrals, (\ref{eneq}), then becomes

\begin{equation}
\begin{array}{l}
\displaystyle{{{\partial  e_1} \over {\partial  t}} +  {\partial \over
{\partial  x}}  [u_1  (e_1 +  p_1  +  {1  \over  2} \rho_1  q_1^2)]  =
H_1 + \sum_{i=2}^N {G_{1i}}} \\
 \\
~~~~~~~~~~~=\displaystyle{{\bf J} \cdot {\bf E} + \sum_{i=1}^N {[H_i +
\alpha_i \rho_i {\bf q}_i \cdot ({\bf q}_i \wedge {\bf B})]}}.\\
\end{array}
\label{aveneq},
\end{equation}

Equations (\ref{avmomeq}), (\ref{aveneq})  and the continuity equation
for the  neutrals, (\ref{coneq}) with $i  = 1$, are  just the ordinary
gas dynamic  equations with source terms.  They can be  written in the
form

\begin{equation}
{{\partial {\bf  Q}} \over {\partial  t}} + {{\partial {\bf  F}} \over
{\partial x}} = {\bf S},
\label{conseq}
\end{equation}

\noindent
where

\[
{\bf Q} = (\rho_1, \rho_1 {\bf q}_1, e_1),
\]

\noindent
is a vector of conserved variables,  ${\bf F}$ is the vector of fluxes
and ${\bf S}$ is  the vector of source terms. We have  ${\bf F} = {\bf
F}({\bf Q})$, but the source  term depends upon the magnetic field and
the state of the other fluids. We therefore write

\[
{\bf S} = {\bf S}({\bf Q}, {\bf V})
\]

\noindent
where ${\bf  V}$ is a  vector representing ${\bf B}$,  $\rho_i$, ${\bf
q}_i$ etc.

\subsection{Numerical Scheme for the Neutral Fluid}

Equations  (\ref{conseq}) are  solved using  the second  order Godunov
scheme described  in Falle (1991).   This is a conservative  scheme in
which the numerical  solution at time $t_n$ in  the cell with $(j-1/2)
\Delta  x \le x  \le (j+1/2)  \Delta x$  is defined  to be  the volume
average

\[
{\bf   Q}^n_{j}   =    {1   \over   {\Delta   x}}   \int_{(j-1/2)\Delta
x}^{(j+1/2)\Delta x} {{\bf U}(t_n,x)~dx}.
\]

\noindent
Integrating  (\ref{conseq}) over  the  $j$th cell  and  from $t_n$  to
$t_{n+1} = t_n + \Delta t_n$, gives

\begin{equation}
\begin{array}{l}
\displaystyle{{1 \over {\Delta t_n}}({\bf  Q}^{n+1}_j - {\bf Q}^n_j) +
{1 \over \Delta x}({\bf F}^{n+1/2}_{j+1/2} - {\bf F}^{n+1/2}_{j-1/2})}
\\
 \\
~~~~~~~~~~~~~~ = {\bf S}^{n+1/2}_j,
\end{array}
\label{numconseq}
\end{equation}

\noindent
where ${\bf F}^{n+1/2}_{j\pm1/2}$ are  the time averages of the fluxes
at the cell  edges and ${\bf S}^{n+1/2}_j$ is the  time average of the
integral of the source term over the cell.

We start by using equation  (\ref{numconseq}) to compute a first order
approximation, ${\bf Q}^{n+1/2}_j$, to  the solution at the half time,
$t_{n+1/2} = t_n  + \Delta t_n/2$. In this step  the fluxes and source
term are approximated by

\[
{\bf    F}^{n+1/2}_{j+1/2}    =    {\bf   F}_*({\bf    Q}^n_j,    {\bf
Q}^n_{j+1}),~~~~{\bf S}^{n+1/2}_j = {\bf S}({\bf Q}^n_j, {\bf V}^n_j),
\]

\noindent
where ${\bf  F}_*({\bf Q}_L, {\bf Q}_R)$  is the flux  in the resolved
state for a  gas dynamic Riemann problem for which  the left and right
states are ${\bf Q}_L$ and ${\bf Q}_R$.

To make the  scheme second order in space, we use  the solution at the
half time to construct an average gradient of the primitive variables,
${\bf P} = (\rho_1, {\bf q}_1, p_1)$, in the $i$th cell

\[
{\left( {{{\partial {\bf P}} \over {\partial x}}} \right)}_j^{n+1/2} =
{1  \over {\Delta  x}} av({\bf  P}^{n+1/2}_{j+1} -  {\bf P}^{n+1/2}_j,
{\bf P}^{n+1/2}_j - {\bf P}^{n+1/2}_{j-1}),
\]

\noindent
where the averaging function is

\[
av(a,b) = \left\{{
\begin{array}{ll}
0 & {\rm  if~} ab < 0,\\  \displaystyle{{{a^2 b + a b^2}  \over {a^2 +
b^2}}} & {\rm otherwise}.\\
\end{array}
}\right.
\]

\noindent
The averaging  function acts as  a non-linear switch that  reduces the
scheme to first  order in regions with large  second derivatives, such
as shocks etc (see e.g. van Leer 1977).

The solution at  $t_{n+1} = t_n + \Delta t_n$  is then calculated from
(\ref{numconseq}) with

\[
{\bf F}^{n+1/2}_{j+1/2}  = {\bf F}_*[{\bf Q}({\bf  P}_L), {\bf Q}({\bf
P}_R)],
\]

\noindent
where

\[
\begin{array}{lcl}
{\bf P}_L & = & \displaystyle{{\bf  P}^{n+1/2}_j + {1 \over 2} \Delta x
{\left( {{{\partial {\bf P}} \over {\partial x}}} \right)}_j^{n+1/2}} \\
 & & \\
{\bf  P}_R &  = &  \displaystyle{{\bf P}^{n+1/2}_{j+1}  - {1  \over 2}
\Delta   x  {\left(   {{{\partial  {\bf   P}}  \over   {\partial  x}}}
\right)}_{j+1}^{n+1/2}},
\end{array}
\]

\noindent
and the source term given by

\[
{\bf S}^{n+1/2}_j = {\bf S}({\bf Q}^{n+1/2}_j,{\bf V}^{n+1/2}_j).
\]

This is obviously not a complete  scheme since we have yet to devise a
method of advancing the variables, ${\bf V}$, describing the field and
charged fluids.

\subsection{Equations for the Charged Fluids and Fields}
  
The reduced momentum equations,  (\ref{redmom}) and the expression for
the current, (\ref{curr}) can be  solved for the electric field. It is
standard practice to write the result in the form

\begin{equation}
{\bf E} = - {\bf q} \wedge {\bf B} + r_0 {{({\bf J} \cdot {\bf B}) {\bf
B}} \over B^2} + r_1 {{{\bf J}  \wedge {\bf B}} \over B} - r_2 {{({\bf
J} \wedge {\bf B}) \wedge {\bf B}} \over B^2}
\label{efield}
\end{equation}

\noindent
(Cowling  1957;  Nakano  \&  Umebayashi  1986).   Here  $r_0$  is  the
resistivity along the  field, $r_1$ is the Hall  resistivity and $r_2$
is the ambi-polar resistivity.  If we define the conductivity parallel
to the field,

\[
\sigma_0 = {1 \over B} \sum_{i=2}^N {\alpha_i \rho_i \beta_i}
\]

\noindent
the Hall conductivity,

\[
\sigma_1 =  {1 \over B}  \sum_{i=2}^N {{{\alpha_i \rho_i} \over  {(1 +
\beta_i^2)}}}
\]

\noindent
and the Pedersen conductivity,

\[
\sigma_2 = {1 \over  B} \sum_{i=2}^N {{{\alpha_i \rho_i \beta_i} \over
{(1 + \beta_i^2)}}}
\]

\noindent
then the resistivities are

\[
r_0     =    1/\sigma_0,~~~r_1    =     \sigma_1/(\sigma_1^2    +
\sigma_2^2),~~~r_2 = \sigma_2/(\sigma_1^2 + \sigma_2^2)
\]

Substituting  (\ref{efield}) into  the Faraday  equations (\ref{Fara})
and using  Ampere's law (\ref{Amp})  to eliminate ${\bf J}$  gives the
induction equation

\begin{equation}
{{\partial {\bf B}} \over {\partial  t}} + {{\partial {\bf M}} \over
{\partial  x}}  = {\partial  \over  {\partial  x}} {\bf R}  {{\partial
{\bf B}} \over {\partial x}},
\label{indeq}
\end{equation}

\noindent
where the hyperbolic flux is

\[
{\bf M} = (0, uB_y - vB_x, uB_z - wB_x),
\]

\noindent
and the resistance matrix is  

\[
{\bf R} = r_0 {\bf R}_0 + r_1 {\bf R}_1 + r_2 {\bf R}_2,
\]

\noindent
with

\[
{\bf R}_0 = \left({
\begin{array}{cc}
\displaystyle{{{B_z^2} \over  B^2}} & \displaystyle{-{{B_y  B_z} \over
B^2}} \\
 & \\
\displaystyle{-  {{B_y B_z} \over  B^2}} &  \displaystyle{{B_y^2 \over
B^2}} \\
\end{array}
} \right),
\]

\[
{\bf R}_1 = \left({
\begin{array}{cc}
0 & \displaystyle{{B_x \over B}}\\
 & \\
\displaystyle{- {B_x \over B}} & 0 \\
\end{array}
} \right),
\]

\[
{\bf R}_2 = \left({
\begin{array}{cc}
\displaystyle{1 - {{B_z^2} \over B^2}} & \displaystyle{{{B_y
B_z} \over B^2}} \\
 & \\
\displaystyle{{{B_y B_z} \over B^2}}  & \displaystyle{1 - {B_y^2 \over
B^2}}\\
\end{array}
} \right).
\]

Equation (\ref{indeq}) can used to advance the field, but, as we shall
see, an explicit approximation to this is very inefficient if the Hall
term is much larger than the ambi-polar diffusion term.

\subsection{Numerical Stability}

The presence of the diffusive terms on the right hand side of equation
(\ref{indeq})  has  some  important  implications  for  any  numerical
scheme, the most obvious of which is that the stable time step for any
explicit  scheme  will be  proportional  to  the  square of  the  mesh
spacing. What  is less obvious  is that the  Hall term places  an even
more severe restriction on the stable time step for explicit schemes.

In  order to see  this, suppose  that the  diffusive flux  in equation
(\ref{indeq}) is much larger than the hyperbolic flux.  Since $r_0$ is
much smaller  than the  other resistivities provided  that there  is a
significant  charge density of  particles with  large $\beta$,  we can
also ignore  the term $r_0 {\bf R_0}$ in the resistance matrix.

If we now linearise (\ref{indeq}) about a state in which $B_z = 0$ and
the field makes an angle $\theta$ with the x-axis, then

\[
{\bf R} = \left({
\begin{array}{cc}
r_2 &  r_1 \cos \theta\\
 & \\
-r_1 \cos \theta &  r_2\cos^2 \theta\\
\end{array}
} \right).
\]

Now let ${\bf B}^n_j$ be  a numerical approximation to ${\bf B}(t_n, j
\Delta x)$ and consider the obvious scheme

\begin{equation}
\begin{array}{l}
\displaystyle{{1 \over {\Delta t}} ({\bf B}^{n+1}_j - {\bf B}^n_j)
=} \\
 \\
~~~~\displaystyle{ {1  \over {\Delta  x^2}}  {\bf  R}  [(1 -  \mu)
({\bf B}^n_{j+1} - 2 {\bf B}^n_j + {\bf B}^n_{j-1})} \\
 \\
~~~~~~~~~~~~~~~\displaystyle{  +  \mu ({\bf  B}^{n+1}_{j+1}  - 2  {\bf
B}^{n+1}_j + {\bf B}^{n+1}_{j-1})]},
\end{array}
\label{numindeq1}
\end{equation}

\noindent
where $0 \le \mu \le 1$. This scheme is purely explicit for $\mu = 0$,
purely implicit for  $\mu = 1$ and time  centred (Crank-Nicolson) for
$\mu = 1/2$. For a numerical wave of the form

\[
{\bf B}^n_j = {\bf B}^n exp(i \omega j),
\]

\noindent
we get

\[
{\bf B}^{n+1} = {\bf A} {\bf B}^n,
\]

\noindent
where the amplification matrix, ${\bf A}$ is

\[
{\bf A}  = [{\bf I} +  \nu \mu {\bf R}]^{-1}[{\bf  I} - \nu  (1 - \mu)
{\bf R}],
\]

\noindent
with

\[
\nu = {{2 \Delta t (1 -  \cos \omega)} \over {\Delta x^2}}.
\]

\noindent
and  ${\bf I}$ the  $2 \times  2$ identity  matrix.  Clearly  the most
restrictive condition on  the timestep is for the  $\pm 1$ mode, $\cos
\omega = -1$, for which

\begin{equation}
\nu = {{4 \Delta t} \over {\Delta x^2}}.
\label{nu}
\end{equation}

\subsubsection{Explicit Scheme ($\mu = 0$)}

In this case the eigenvalues of ${\bf A}$ are given by

\[
\begin{array}{l}
{\lambda}^{2}+   \left(  \nu\,r_2\,\cos^2   \theta-2+\nu\,r_2  \right)
\lambda \\
~~~+1-\nu\,r_2\,\cos^2    \theta-\nu\,r_2+{\nu}^{    2}{r_1}^{2}\cos^2
\theta+{\nu}^{2}{r_2}^{2}\cos^2 \theta = 0.
\end{array}
\]

\noindent
This has real roots if

\[
r_1 \cos \theta \le {1 \over 2} r_2 \sin^2 \theta,
\]

\noindent
in which case stability requires

\[
\nu \le {4 \over {[r_2 \cos^2 \theta + r_2 + \surd(r_2^2 \sin^4 \theta
- 4 r_1^2 \cos^2 \theta)]}}.
\]

This increases with $r_1 \cos \theta$, so that the Hall term increases
the stable  time step,  but when the  eigenvalues of ${\bf  A}$ become
complex, the stability condition is

\[
\nu  \le {{r_2  (1 +  \cos^2 \theta)}  \over {(r_1^2  +  r_2^2) \cos^2
\theta}}.
\]

\noindent
This tells us that the stable time step tends to zero as the Hall term
becomes large  compared to ambi-polar  diffusion, which means  that an
explicit scheme is very inefficient  in such cases. This would seem to
be the explanation for the  severe restrictions on the stable timestep
experienced by Hollerbach \& R\"udiger (2002) in their calculations of
the Hall effect in neutron stars.

\subsubsection{Crank-Nicolson Scheme ($\mu = 1/2$)}

The eigenvalues of ${\bf A}$ are now given by

\[
\begin{array}{l}
(2 \nu  r_2 \cos^2 \theta+2 \nu r_2+4+\nu^2  r_1^2 \cos^2 \theta+\nu^2
r_2^2 \cos^2  \theta) \lambda^2\\
~~+(2  \nu^2  r_1^2  \cos^2   \theta+2  \nu^2  r_2^2  \cos^2  \theta-8)
\lambda\\
~~~~+4+\nu^2  r_1^2  \cos^2  \theta-2  \nu  r_2  \cos^2  \theta-2  \nu
r_2+\nu^2 r_2^2 \cos^2 \theta = 0,
\end{array}
\]

\noindent
whose roots are

\[
\lambda  =  {{4  -\nu^2  \cos^2  \theta  (r_1^2 +  r_2^2)  \pm  2  \nu
\surd(r_2^2 \sin^4 \theta  - 4 r_1^2 \cos^2 \theta)} \over  {4 + 2 \nu
r_2 (1 + \cos^2 \theta) + \nu^2 \cos^2 \theta (r_1^2 + r_2^2)}}.
\]

\noindent
As we would expect, $|\lambda| \le  1$ for all $\nu$, so the scheme is
unconditionally stable.

One  might wonder  whether there  is some  other explicit  scheme with
better  stability properties.  That this  is unlikely  can be  seen by
looking at the nature of the  equation when the Hall term is dominant.
We then get a dispersive wave equation for whistler waves with

\[
\omega^2 = {{r_1^2 B_x^2} \over B^2} k^4,
\]

\noindent
so that  both the phase and  group velocities tend to  infinity as the
wavenumber tends  to infinity.  This  not only explains why  the first
order explicit approximation, (\ref  {numindeq1}), to the Hall term is
unstable,  but  also  suggests   that  there  is  no  stable  explicit
approximation. For example, the second order explicit scheme described
by Sano \& Stone (2002) is unconditionally unstable when the Hall term
is dominant, although this was  not apparent in their test calculation
because the numerical  resolution was low enough for  the scheme to be
stabilized by the hyperbolic term.

Ciolek  \& Roberge  (2002) do  not  use (\ref{indeq})  to advance  the
field, instead  they write ${\bf E}$  in terms of  the velocity, ${\bf
q}_3$ of the  ions

\[
{\bf E}= - {\bf q}_3 \wedge {\bf B},
\]

\noindent
which  is valid  since  $\beta_3  \gg1$.  They  then  use the  Faraday
equation, (\ref{Fara}), to obtain an evolution equation for ${\bf B}$.
This has the disadvantage that, since there is no simple way of making
such a  scheme implicit, it  requires a very  small time step  at high
resolution,  particularly when  the  Hall term  is  large compared  to
ambi-polar diffusion.

\subsection{Numerical Scheme for the Charged Fluids and Fields}

In order to  obtain the state of the charged  fluids and fields, ${\bf
V}_{n+1/2j}$ at  the half  time, we first  advance the  magnetic field
with the first order scheme

\begin{equation}
\begin{array}{lcl}
\displaystyle{{1 \over {\Delta t}} ({\bf B}^{n+1/2}_j - {\bf B}^n_j) +
{1 \over {\Delta x}} ({\bf M}^n_{j+1/2} - {\bf M}^n_{j-1/2})} & = & \\
 & & \\
\multicolumn{3}{r}{\displaystyle{ {1  \over {\Delta x^2}}  {\bf R}^n_j
 ({\bf    B}^{n+1/2}_{j+1}   -    2   {\bf    B}^{n+1/2}_j    +   {\bf
 B}^{n+1/2}_{j-1})}}, \\
\end{array}
\label{numindeq2}
\end{equation}

\noindent
where ${\bf R}^n_j = {\bf R}({\bf V}^n_j)$.  Note that the term on the
right hand side are calculated implicitly, whereas the hyperbolic term
is  calculated explicitly. This  is a  block tridiagonal  equation for
${\bf B}^{n+1/2}$ with the blocks  consisting of $2 \times 2$ matrices
and can readily be solved by Gaussian elimination.

It would  be nice to use  a Riemann problem to  compute the hyperbolic
flux, ${\bf  M}$, but this  is not possible  because we would  have to
solve a Riemann  problem in which the magnetic field  does not exert a
force on  the gas. In general  the solution to  such problems contains
discontinuities  in the tangential  velocities which  are incompatible
with the  induction equation. We therefore  have to be  content with a
centred approximation to the hyperbolic flux

\[
{\bf M}^n_{j+1/2} = {1 \over 2} ({\bf M}^n_{j+1} + {\bf M}^n_j).
\]

\noindent
This  is perfectly  satisfactory as  long as  the resistive  terms and
numerical  resolution are  such  as that  the  magnetic field  appears
continuous on the grid.

The densities of the charged fluids can be calculated from an explicit
upwind approximation to the  continuity equations and the current from
a centred  approximation to the  Ampere's law, (\ref{Amp}),  using the
field at  the half time. Given ${\bf  J}$, ${\bf B}$ and  the state of
the  neutral  fluid   at  the  half  time,  we   can  calculate  ${\bf
V}^{n+1/2}_j$ from the reduced momentum equations, (\ref{redmom}), the
expression  for  the current,  (\ref{curr}),  and  the reduced  energy
equations,  (\ref{reneqi}).   Note that  these  equations  have to  be
solved by iteration if  the interaction coefficients, $K_{1i}$, depend
upon the velocities, but since the solution at the old time provides a
very good initial guess, this is not expensive.

${\bf V}^{n+1/2}_j$ can  then be used to calculate  the source term at
the half time so that the neutral solution can be advanced to the full
time  using  (\ref{numconseq})  with  the second  order  fluxes.   The
densities  of the  charged fluids  can  also be  advanced using  ${\bf
V}^{n+1/2}_j$  in a  second  order approximation  to their  continuity
equations. As we  shall see, this algorithm is  well behaved even when
some of the charged species have very large Hall parameters.

The  magnetic field  is advanced  explicitly  to the  full time  using
fluxes computed from the solution at the half time

\[
\begin{array}{lcl}
\displaystyle{{1 \over  {\Delta t}} ({\bf B}^{n+1}_j -  {\bf B}^n_j) +
{1    \over   {\Delta    x}}   ({\bf    M}^{n+1/2}_{j+1/2}    -   {\bf
M}^{n+1/2}_{j-1/2})} & = & \\
 & & \\
\multicolumn{3}{r}{\displaystyle{   {1   \over   {\Delta  x^2}}   {\bf
R}^{n+1/2}_j  ({\bf  B}^{n+1/2}_{j+1} -  2  {\bf  B}^{n+1/2}_j +  {\bf
B}^{n+1/2}_{j-1})}}. \\
\end{array}
\]

\noindent
Although this  is explicit, the fact  that the field at  the half time
has been  calculated implicitly ensures the  same stability properties
as  a Crank-Nicolson.   Finally the  neutral solution,  charged fluid
densities  and magnetic  field can  be used  to calculate  the charged
fluid velocities and temperatures at the  full time in the same way as
for ${\bf V}^{n+1/2}_j$.

\section{Test Calculations}

\begin{table*}
\centering
\begin{minipage}{18cm}
\caption{Parameters for the Test Calculations}
\centering
\begin{tabular}{llllll}
Case A & & & & & \\
Upstream & $\rho =  1$ & ${\bf q} = (-1.7510, 0, 0)$  & ${\bf B} = (1,
0.6, 0)$ & $\rho_2 = 5~10^{-8}$  & $\rho_3 = 10^{-3}$\\

Downstream &  $\rho = 1.7942$  & ${\bf q}  = (-0.9759, -0.6561,  0)$ &
${\bf B}  = (1, 1.74885, 0)$  & $\rho_2 = 8.9712~10^{-8}$  & $\rho_3 =
1.7942~10^{-3}$ \\
& $\alpha_2 =  - 2~10^{12}$ & $\alpha_3 = 10^8$ &  $K_{12} = 4~10^5$ &
 $K_{13} = 2~10^4$ & $a = 0.1$ \\
Case B & & & & & \\
Upstream & \multicolumn{5}{l}{as for case A} \\
Downstream & \multicolumn{5}{l}{as for case A} \\
 & $\alpha_2 =  - 2~10^{9}$ & $\alpha_3 = 10^5$ &  $K_{12} = 4~10^2$ &
$K_{13} = 5~10^5$ & $a = 0.1$ \\
Case C & & & & & \\
Upstream & $\rho =  1$ & ${\bf q} = (-6.7202, 0, 0)$  & ${\bf B} = (1,
0.6, 0)$ & $\rho_2 = 5~10^{-8}$  & $\rho_3 = 10^{-3}$\\
Downstream &  $\rho = 10.421$  & ${\bf q}  = (-0.6449, -1.0934,  0)$ &
${\bf B}  = (1, 7.9481,  0)$ & $\rho_2  = 5.2104~10^{-7}$ &  $\rho_3 =
1.0421~10^{-2}$ \\
 & $\alpha_2 = - 2~10^{12}$ &  $\alpha_3 = 10^8$ & $K_{12} = 4~10^5$ &
$K_{13} = 2~10^4$ & $a = 1$ \\
\end{tabular}
\label{table}
\end{minipage}
\end{table*}

In  order to  test  the code  described  in the  previous section,  we
compare the results for steady  shocks with those obtained by a direct
solution of the steady  equations. The numerical solution was obtained
by imposing the  upstream and downstream states at  the right and left
boundaries of  the domain  with a discontinuity  in the middle  of the
domain and then integrating until the steady state is reached.

\subsection{Steady Solutions}

In order  to calculate the  steady solution, we start  by transforming
the induction equation,  (\ref{indeq}), to a frame in  which the shock
is steady and then setting the time derivative to zero. This gives

\[
{{d {\bf M}} \over {d x}} = {d \over {d x}} {\bf R} {{d {\bf B}} \over
{d x}},
\]

\noindent
which can be integrated to give

\begin{equation}
{\bf M} - {\bf M}_L = {\bf M} - {\bf M}_R = {\bf R} {{d {\bf B}} \over
{d x}},
\label{stindeq}
\end{equation}

\noindent
where the suffices $L$, $R$, denote the upstream and downstream states
respectively. The shock relations ensure  that ${\bf M}_L = {\bf M}_R$
so that  the upstream and downstream  states are fixed  points of this
equation.

We confine ourselves to the  case where the neutral gas is isothermal,
so that  the steady versions  of the neutral  equations can be  all be
integrated to give

\[
\begin{array}{l}
\rho u = Q,~~~~\rho u^2 + a^2  \rho + {1 \over 2} B^2 = P_x \\
\rho u v - B_x B_y = P_y,~~~~\rho u w - B_x B_z = P_z \\
\end{array}
\]

\noindent
where  $Q$,  $P_x$, $P_y$  and  $P_z$ are  constants  and  $a$ is  the
isothermal  neutral  sound speed.  Here  we  have  used Ampere's  law,
(\ref{Amp}), to write ${\bf J}$ in  terms of ${\bf B}$ in the momentum
equation. These equations can be solved  to give $\rho$ and ${\bf q} =
(u, v, w)$ as functions of ${\bf B}$.

\begin{figure} 
\begin{center} 
\epsffile[190 210 409 617]{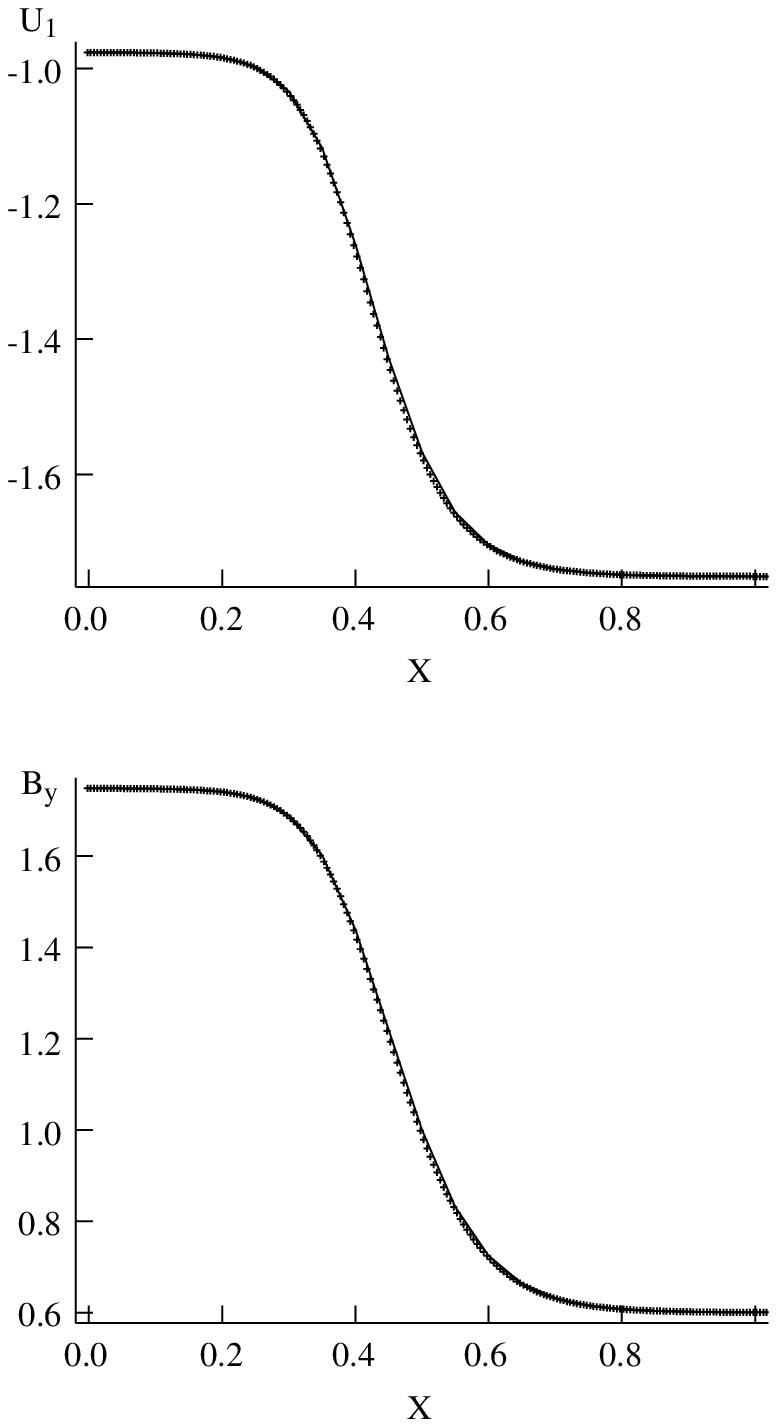}
\caption{Neutral x velocity and y component of magnetic field for case
A with $\Delta x = 5~10^{-3}$.  The line is the solution to the steady
equations and the markers are the time dependent numerical solution.}
\end{center}
\end{figure}

\begin{figure} 
\begin{center} 
\epsffile[190 294 409 500]{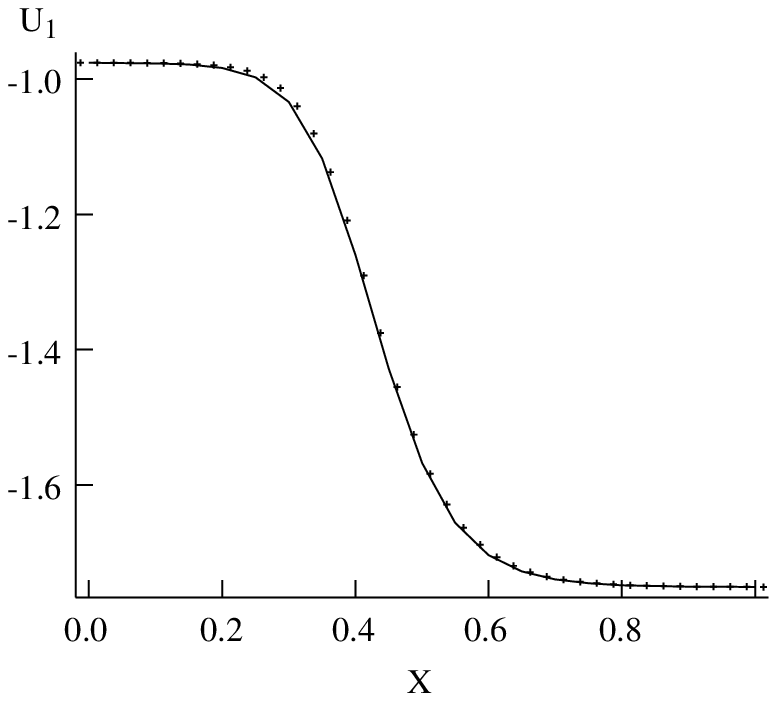}
\caption{Neutral x velocity for case  A with $\Delta x = 2.5~10^{-2}$.
The line is  the solution to the steady equations  and the markers are
the time dependent numerical solution.}
\end{center}
\end{figure}

\begin{figure} 
\begin{center} 
\epsffile[190 124 409 747]{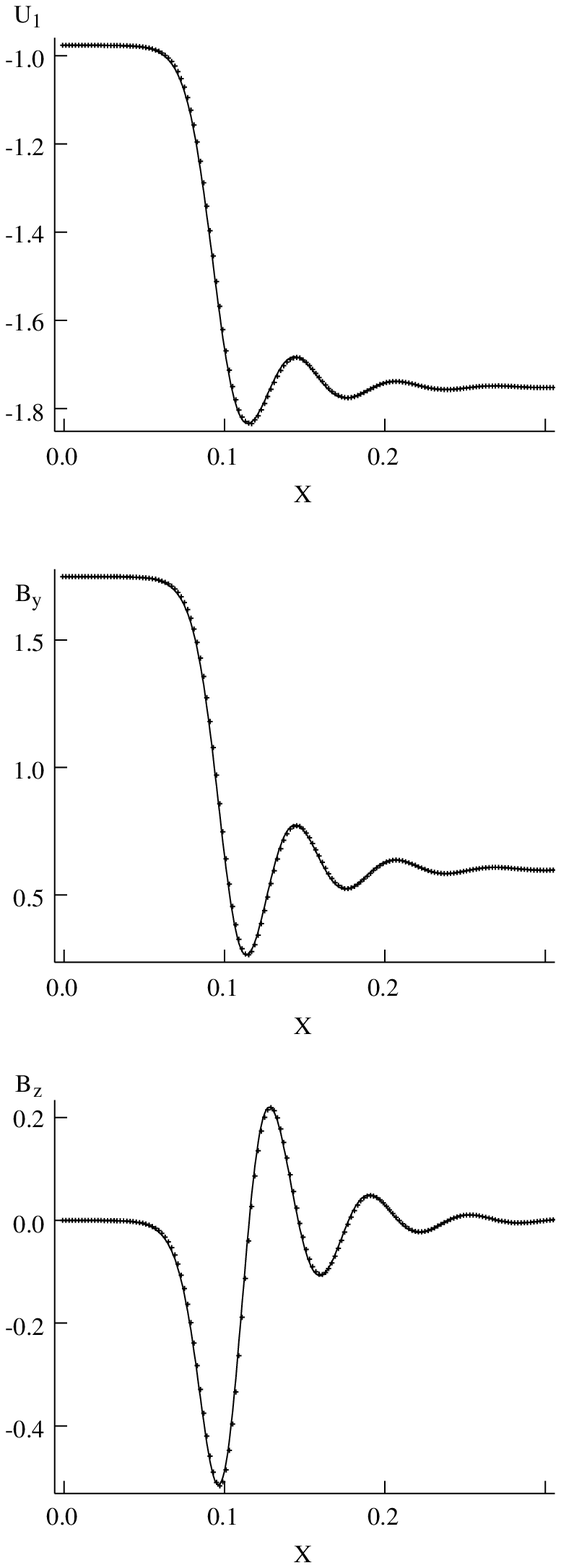}
\caption{Neutral x  velocity and y,z components of  magnetic field for
case B $\Delta x = 2~10^{-3}$.  The line is the solution to the steady
equations and the markers are the time dependent numerical solution.}
\end{center}
\end{figure}

\begin{figure} 
\begin{center} 
\epsffile[190 294 409 500]{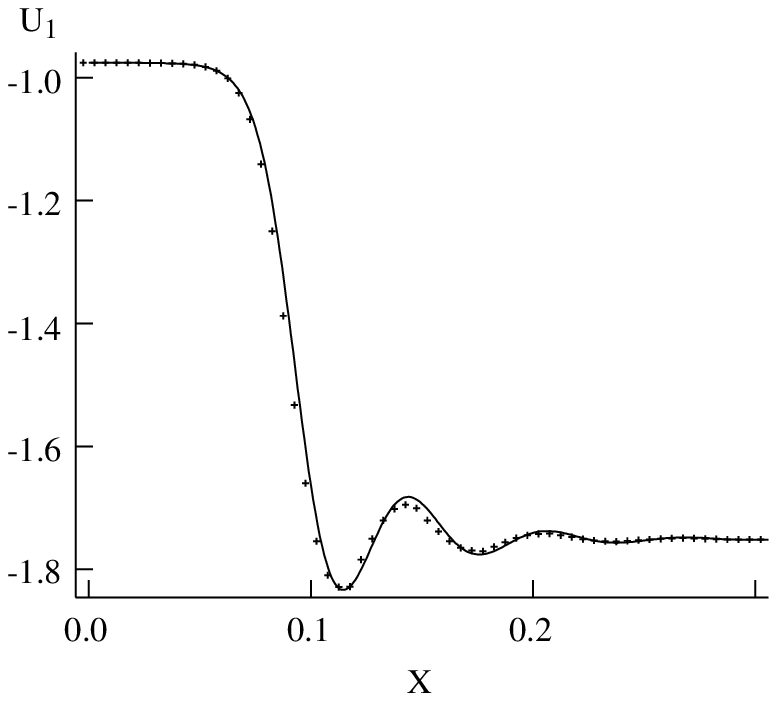}
\caption{Neutral x  velocity for case  B with $\Delta x  = 5~10^{-3}$.
The line is  the solution to the steady equations  and the markers are
the time dependent numerical solution.}
\end{center}
\end{figure}

\begin{figure} 
\begin{center} 
\epsffile[190 124 409 747]{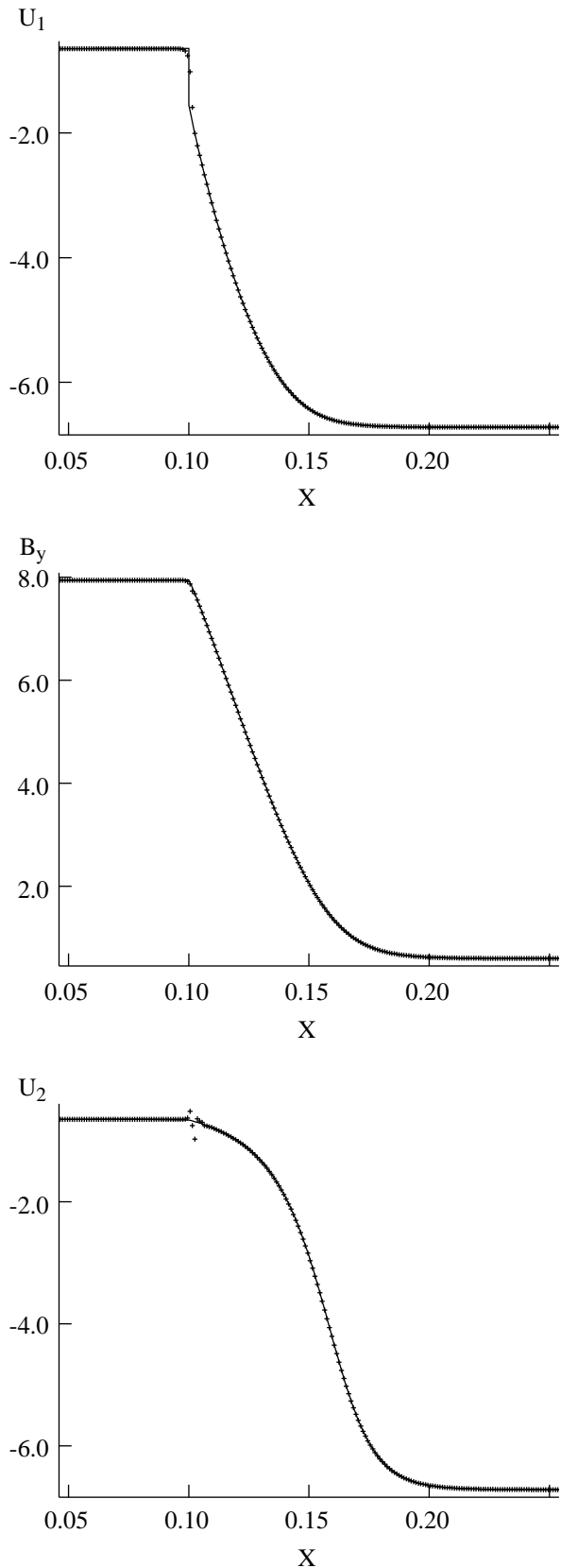}
\caption{Neutral x velocity, y component of magnetic field and fluid 2
x velocity  for case C with $\Delta  x = 10^{-3}$.  The  line is the
solution  to  the  steady  equations  and the  markers  are  the  time
dependent numerical solution.}
\end{center}
\end{figure}

\begin{figure} 
\begin{center} 
\epsffile[190 294 409 500]{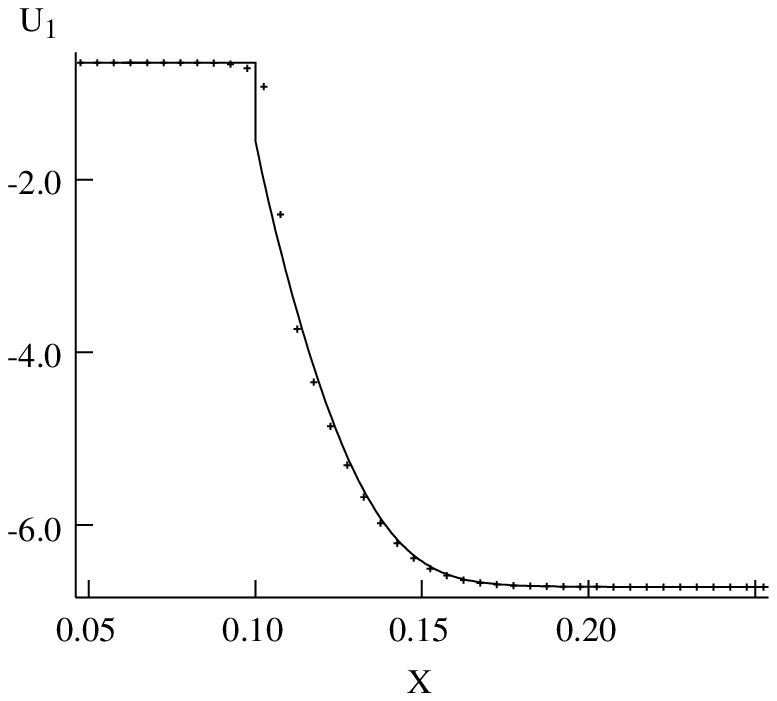}
\caption{Neutral x  velocity for case  C with $\Delta x  = 5~10^{-3}$.
The line is  the solution to the steady equations  and the markers are
the time dependent numerical solution.}
\end{center}
\end{figure}

The charged fluid densities are given by

\begin{equation}
\rho_i u_i = Q_i,
\label{chcon}
\end{equation}

\noindent
where the $Q_i$ are constants.  The reduced momentum equations for the
charged  fluids,  (\ref{redmom}), give  us  $3N-3$  equations for  the
$3N-3$  components  of  the  charged  fluid  velocities  and  the  $3$
components of  the electric  field.  In the  steady case,  the Faraday
equation, (\ref{Fara}), tells us that $E_y$ and $E_z$ are constant and
equal to  their values in  the upstream state,  so that only  $E_x$ is
unknown. We  therefore have $3N-2$ unknowns and  $3N-3$ equations. The
remaining equation  can be obtained from (\ref{chcon})  and the charge
neutrality condition, (\ref{charneut}).

Given ${\bf B}$, we can therefore calculate the neutral velocities and
the charged densities  and hence ${\bf M}$ and  the resistance matrix,
${\bf  R}$.   We  shall  confine  ourselves to  cases  for  which  the
downstream state is  a saddle point and the upstream  state is a sink,
in  which case  the steady  solution  can be  obtained by  integrating
(\ref{stindeq})  from  downstream  to  upstream.   Since  the  gas  is
isothermal,  any  sub-shock must  be  at  the  downstream end  of  the
structure. It  is therefore a simple  matter to insert  a subshock and
then to  integrate from its  upstream side towards the  upstream state
(see case C).

\subsection{Case A (Negligible Hall Effect)}

This has two charged fluids, both of which have $\beta \gg 1$, so that
the Hall resistivity  is neglible.  The parameters are  given in table
\ref{table}.   This  corresponds to  an  oblique  fast  shock with  an
upstream Mach number relative to the fast speed $M_f = 1.5$ and a very
small  neutral pressure. The  charge to  mass ratios,  $\alpha_i$, and
collision   coefficients,   $K_{1,i}$  are   such   that  $\beta_2   =
-5.831~10^6$, $\beta_3 = -5.831~10^3$  in the upstream state, which is
appropriate for electrons  and ions in material with  $n_H = 10^6$ and
$B = 10^{-3}$  G (Wardle 1998). This gives $r_0  = 2~10^{-12}$, $r_1 =
1.16~10^{-5}$ and $r_2  = 0.068$, so that the  Hall term is negligible
compared with ambi-polar diffusion.

From Fig 1, which shows the  neutral x velocity and the y component of
the magnetic  field, it  is clear that  the agreement between  the two
solutions  is excellent  at high  resolution. Even  at the  much lower
resolution shown in Fig 2, the errors are very small. 

In order  to make this  quantitative, we define  the $L_1$ error  in a
primitive variable, $P$, by

\begin{equation}
\epsilon_P = { 1 \over {(j_1 - j_0)}} \sum_{j=j_0}^{j_1} {\left[{P_j -
P_s(j \Delta x)} \right]},
\label{err}
\end{equation}

\noindent
where $P_j$ is the numerical solution  in cell $j$ and $P_s(x)$ is the
solution calculated  from the steady  equations. Here $j_0$  and $j_1$
are such that the region $x_0 =  j_0 \Delta x$ to $x_1 = j_1 \Delta x$
covers the shock structure, but  does not include much of the upstream
and downstream  uniform regions.   Since the numerical  shock position
depends upon the evolution from the initial data, we impose a shift in
x on the steady solution so as to minimize the error.

For case A, with $x_0 = -0.06$, $x_1 = 0.94$, the error in the neutral
x velocity is  $\epsilon_{u_1} = 5.68~10^{-5}$ for $\Delta  x = 0.005$
and  $\epsilon_{u_1} =  1.26~10^{-3}$ for  $\Delta x  =  0.025$.  This
corresponds  to $\epsilon_{u_1} \propto  \Delta x^{1.92}$,  i.e.  very
nearly  second  order  convergence.   For  $\Delta x  =  0.0125$,  the
corresponding error is $3.52~10^{-4}$, which, when compared to $\Delta
x = 0.005$, gives a convergence rate of $\epsilon_{u_1} \propto \Delta
x ^{1.96}$.  This indicates  that the asymptotic convergence is indeed
second order, as we would expect for a smooth solution.

\subsection{Case B (Large  Hall Effect)}

This case also has two  charged fluids with $\beta_2 = -5.831~10^6$ as
in case A, but with $\beta_3 = 0.2332$.  This gives $r_0 = 2~10^{-9}$,
$r_1 =  0.0116$ and $r_2 =  0.00272$, so that the  Hall term dominates
the  ambi-polar  term.   In  all  other  respects,  the  upstream  and
downstream states are for Case A.

This calculation is more demanding than case A since the fact that the
Hall term is dispersive means that  the upstream state is now a stable
spiral  instead  of  a  stable  node. The  shock  structure  therefore
contains large oscillations  whose wavelength is significantly smaller
than the total width of the shock structure. Nevertheless, Fig 3 shows
that  at  sufficiently  high  resolution, the  agreement  between  the
solutions is excellent  and Fig 4 shows that  it is still satisfactory
even when there are only $\sim 50$ mesh points in the shock structure.
Note that, as  we would expect, there is a  significant z component of
the magnetic field in this case.

In this case, with $x_0 = 0$,  $x_1 = 0.4$, the error in the neutral x
velocity is $\epsilon_{u_1} = 3.79~10^{-4}$ for $\Delta x = 0.002$ and
$\epsilon_{u_1} =  2.37~10^{-3}$ for $\Delta  x = 0.005$.   This gives
$\epsilon_{u_1} \propto  \Delta x^{2.0}$, which shows that  at we have
already   reached    second   order   convergence    rate   at   these
resolutions.  Again,  this  is  what  we would  expect  for  a  smooth
solution.   Note,  however,  that  since the  solution  contains  more
structure  than  case  A,  the  errors  are  larger  even  though  the
resolution is higher.

\subsection{Case C (Neutral Sub-shock)}

This is similar to case A except  that the neutral sound speed $a = 1$
and the upstream fast Mach number  $M_f = 5$, which mean that there is
a neutral  sub-shock at  the downstream end  of the structure.   Fig 5
shows that the agreement between the two solutions is again excellent,
except for the finite width  of the sub-shock in the solution computed
with the  time dependent code. There  is also an error  in the charged
fluid x velocities inside the  sub-shock, but this does not affect the
solution  elsewhere. This  error arises  because of  the  exact steady
solution has  a discontinuity in  the electric field at  the sub-shock
and could be reduced by adding  viscosity to increase the width of the
sub-shock.  However, there  is little  point in  doing this  since the
assumption that the inertia of  the charged particles is negligible is
not  valid within the  sub-shock.  Note  that in  this case  also, the
results are still satisfactory at much lower resolution (Fig 6).

Since  the   neutral  x-velocity  is  discontinuous   at  the  neutral
sub-shock, which is  smeared out in the numerical  solution, we expect
the  rate of  convergence to  be first  order.  In  fact, with  $x_0 =
0.065$,  $x_1 =  0.215$, we  get $\epsilon_{u_1}  =  6.88~10^{-3}$ for
$\Delta x = 0.001$ and $\epsilon_{u_1} = 6.98~10^{-2}$ for $\Delta x =
0.005$.  This gives $\epsilon_{u_1} \propto \Delta x^{1.44}$, which is
significantly better than first order.  This merely indicates that, at
this resolution,  the error is not  entirely dominated by  that at the
sub-shock.  If we  increase the resolution to $\Delta  x = 5~10^{-4}$,
we   get  $\epsilon_{u_1}   =  3.94~10^{-3}$   which   corresponds  to
$\epsilon_{u_1} \propto \Delta x^{0.81}$  when compared to $\Delta x =
0.001$.   That this  is somewhat  worse than  first order  is probably
because  rounding  errors  are   becoming  significant  at  such  high
resolution.

\section{Conclusions}

It is clear  that the code described in this paper  is both robust and
accurate and it also has  the advantage that, unlike explicit methods,
the timestep at high numerical  resolution is not restricted by either
the Hall or ambi-polar  terms. Although this is particularly important
when the Hall term dominates  over ambi-polar diffusion, it also gives
a significant increase  in efficiency at high resolution  when this is
not the case. A code that is efficient at high resolution is necessary
if  one   wishes  to  study   multi-fluid  shock  structures   with  a
distribution of  grain sizes and  realistic physics.  The fact  that a
Hall term that is large  compared to ambi-polar diffusion imposes such
a severe  restriction on  the timestep for  explicit schemes  also has
important  implications for  any numerical  calculations  that include
this effect.  It may be  that at relatively low resolution an explicit
scheme  is  stabilized  by   the  numerical  dissipation  due  to  the
hyperbolic terms, but such a scheme will require a very small timestep
if the resolution is sufficient to give accurate results.

Although  we have  only  described a  one  dimensional algorithm,  its
structure  is  such   that  is  a  simple  matter   to  extend  it  to
multi-dimensions. A multi-dimensional version  of this code would have
numerous applications  to the dynamics of molecular  clouds in general
and  star formation  in  particular. For  example,  Wardle (1991)  has
suggested that  the instability that affects C-type  shocks is generic
and is  likely be  important whenever there  there is  both ambi-polar
diffusion and  a dynamically significant magnetic field.   It can also
be applied to accretion discs, such as proto-planetary discs and those
in  dwarf  novae, in  which  the high  density  reduces  the ion  Hall
parameter to  the point where  both ambi-polar diffusion and  the Hall
effect are significant (e.g. Sano \& Stone 2002).

We have not  considered the other source terms  such as the ionization
and recombination, chemical reactions and radiative cooling, but these
terms do  not present any  obvious difficulty. There may  be occasions
when they are stiff, but this merely requires locally implicit methods
rather than  the globally  implicit method that  we have used  for the
induction equation.

\section*{Acknowledgements}
The author  would like to  thank T.~W.~ Hartquist for  suggesting this
problem and for much helpful  advice and criticism.  The final version
has  also  benefitted  from  helpful  suggestions  from  an  anonymous
referee.  The  workstations on  which these calculations  were carried
out were funded by PPARC  and the steady solutions were computed using
Maple.

\end{document}